# Unified Prediction Model for Employability in Indian Higher Education System

[1]Pooja Thakar, [2]Anil Mehta , [1]Manisha

[1,3]Department of Computer Science, Banasthali University, India

[2]School of Business and Commerce, Manipal University Jaipur, India

**Abstract-** Educational Data Mining has become extremely popular among researchers in last decade. Prior effort in this area was only directed towards prediction of academic performance of a student. Very less number of researches are directed towards predicting employability of a student i.e. prediction of students' performance in campus placements at an early stage of enrollment. Furthermore, existing researches on students' employability prediction are not universal in approach and is either based upon only one type of course or University/Institute. Henceforth, is not scalable from one context to another. With the necessity of unification, data of professional technical courses namely Bachelor in Engineering/Technology and Masters in Computer Applications students have been collected from 17 states of India. To deal with such a data, a unified predictive model has been developed and applied on 17 states datasets. The research done in this paper proves that model has universal application and can be applied to various states and institutes pan India with different cultural background and course structure. This paper also explores and proves statistically that there is no significant difference in Indian Education System with respect to states as far as prediction of employability of students is concerned. Model provides a generalized solution for student employability prediction in Indian Scenario.

## 1. Introduction

Economic development of a country is highly dependent on its higher education. It supplies skilled and trained manpower to the different sectors of the economy. Hence the development of the country depends on its quality manpower. It is evermore truer in the case of India as it has the youngest population in the world. The latest report published by India Today on July 13th, 2016 revealed that hardly seven percent of engineering graduates are suitable for core engineering jobs in India (Chakrabarty, 2016). Recent survey and analysis of ASSOCHAM, April 28, 2016, also depicted that most of the students passing out from higher education institutes are not job ready (assocham.org, 2016). These findings have sparked serious concerns about the mismatch between the education system and the needs of the job market. Nothing can be more disruptive for our social cohesion and sustained economic progress than a large army of educated, unemployed youth. To bridge this immense gap the role of the institute is irrefutable. Nowadays, universities/institutes are data rich, but information poor. To stand out in this competitive world, students should be guided well in the beginning of their course. Gained knowledge can give an edge to the students when they present themselves in the market for employment. The need of the hour is to find a mechanism that could find knowledge from such a vast amount of available data and predict employability of students in the very beginning of their course enrollment. Researches revealed that most of the prior effort in the area of education is only directed towards prediction of performance in academic results only. A very little attempt is made to predict students' employability. Furthermore, existing researches on students' employability prediction are either based upon only one type of course or on single university/institute; thus is not scalable from one context to another. There is a dire need of unification.

A unified prediction model (UPM) has been developed that is scalable enough to be applicable in various milieus (Thakar, Mehta, & Manisha, 2017). It has been designed for predicting employability of students at an initial stage of enrollment and works on two major aspects. First is an automated pre-processing method to find the major set of attributes in dataset that affect the prediction (Thakar, Mehta, & Manisha, 2016). Secondly, ensemble model for better prediction by integrating machine learning methods of classification. In this paper Unified Prediction Model is applied on different datasets of 17 states of India. Model resulted in very encouraging results and proves that model has universal application and can be applied to various states and institutes pan India with different cultural background and course structure. Another objective of the current study is to explore whether the significant difference exists with respect to state in Indian Education System to predict the employability of students.

## 2. Related Work

In last decade, educational data mining has aroused the interest of research community and most of the researches are focused to predict the academic performance of students (Sharma, 2011) (Siraj, 2009) (Pumpuang, 2008) (Romero, 2008) (Nghe, 2007) (Saa, 2016) (El Moucary, 2011) in Educational Data Mining. Only some researchers have worked towards predicting employability of students (Mishra, 2016) (Piad, 2016). Few recent studies created employability models (Chen, 2016) (Jantawan, 2013). All of them are either applied on one course, institute or university. Need is to create a model with universal application capabilities. Almost all of the researches emphasized to include all types of parameters such as English aptitude, cognitive skills, logical skills, psychometric parameters, background with academic attributes for better prediction (Rahmat, 2015) (Potgieter, 2013) (Finch, 2013). The majority of universities and institutes conduct training and take various types of aptitude test. Such data can be used for predicting the employability of students at initial stage of their course enrollment.

## 3. Unified Prediction Model (Upm)

Unified Prediction Model (UPM) has been designed by integrating supervised and unsupervised machine learning techniques (clustering and classification) that could predict Students' Employability at early stages of course enrollment (Thakar, Mehta, & Manisha, 2017). The model works in three phases. The first phase implements the concept of automated pre-processing, where raw data is converted to refined data. An innovative approach is used for preprocessing the raw data to the transformed data set. Automated pre-processing reduces dimensionality, find a relevant set of attributes and provide a refined, transformed dataset, which can further be readily used for better classification results. Then it is taken further at second phase for ensemble classification. Instead of choosing one method for classification, voting ensemble method is used for improved classification accuracy. Random Tree, K-Star, Simple Cart and Random Forest are selected as base learners. Thus, it integrates four classifiers to predict student's employability (i. e. Placed or Unplaced in on-campus placement drives). Last third phase generates rules to facilitate decision making. Rules are generated by the simple method of converting trees to rules. Rules generated help in identifying grey areas of students to focus upon. Timely corrective actions can help in improving students' performance in final placement drives and make them industry ready. Unified Prediction Model (UPM) thus deals with the complex dataset, automatically selects a relevant set of attributes from a large pool of attributes and integrates machine learning algorithms to predict students' employability precisely. Model is scalable enough to be applied in any context. Thus, proposes an easy and generalized solution to the problem. To further test the robustness of UPM, it is applied on datasets of 17 states of India. Logical Diagram of UPM is presented in figure 1.

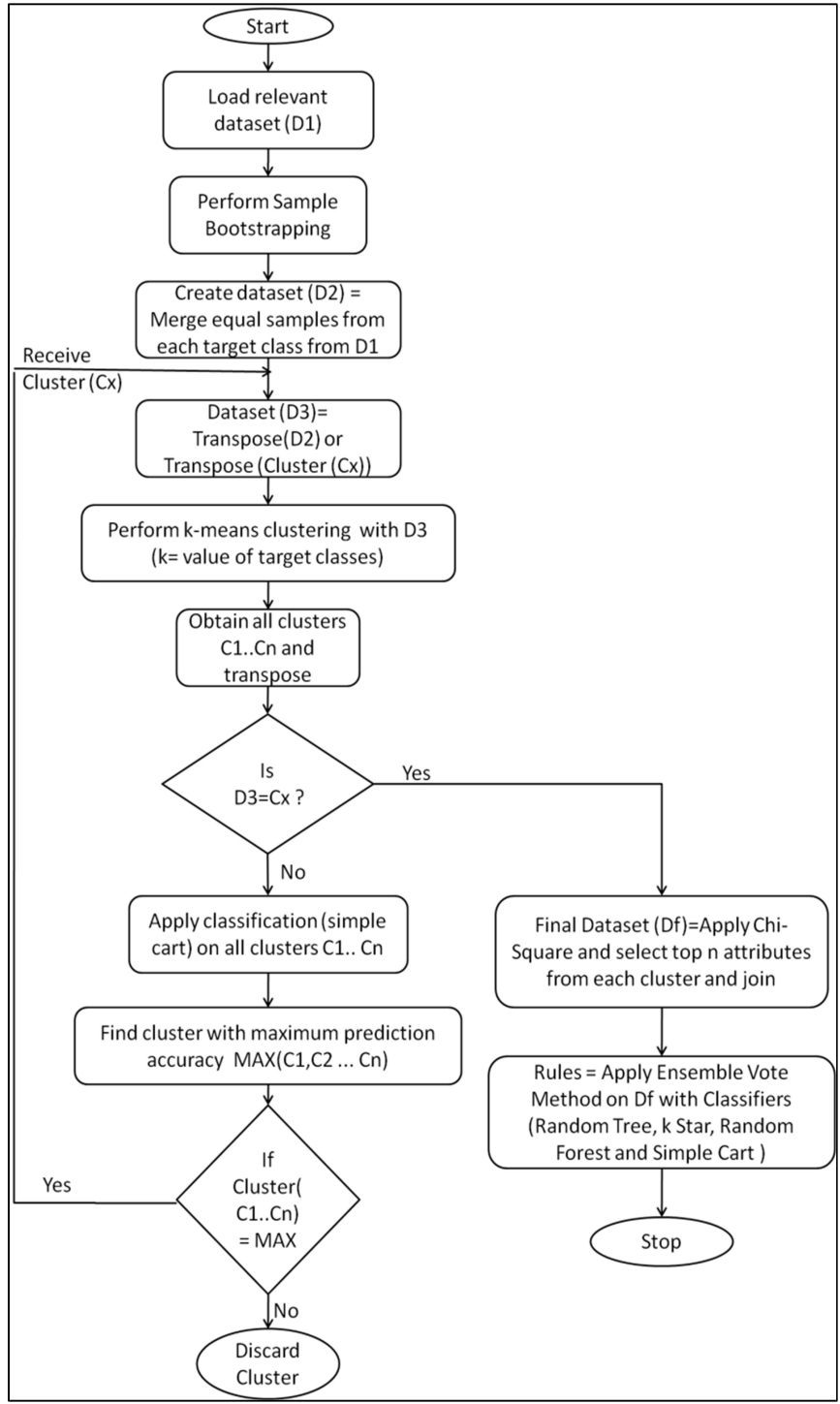

**Figure1. Logical Diagram of Unified Prediction Model**

## 4. Experimental Setup

**A       Datasets**

The dataset was obtained from previous research dataset, which comprised of historical data of Engineering and "Master of Computer Applications" students, collected between the years 2000 and 2016 from various universities and institutes pan India [3]. The dataset thus obtained is further categorized and separated in to 17 datasets representing 17 states of India each having population size more than one million. They are represented in Table 1:

| DATASET | State | Instances in dataset (attributes 150) |
|---------|-------|---------------------------------------|
| DATASET 1 | Andhra Pradesh | 516 |
| DATASET 2 | Bihar | 411 |
| DATASET 3 | Chhattisgarh | 439 |
| DATASET 4 | Delhi | 460 |
| DATASET 5 | Gujarat | 440 |
| DATASET 6 | Haryana | 350 |
| DATASET 7 | Jharkhand | 425 |
| DATASET 8 | Karnataka | 310 |
| DATASET 9 | Kerala | 261 |
| DATASET 10 | Madhya Pradesh | 344 |
| DATASET 11 | Maharashtra | 958 |
| DATASET 12 | Punjab | 453 |
| DATASET 13 | Rajasthan | 178 |
| DATASET 14 | Tamil Nadu | 287 |
| DATASET 15 | Uttar Pradesh | 1192 |
| DATASET 16 | Uttarakhand | 104 |
| DATASET 17 | West Bengal | 32 |

**Table 1Datasets from 17 States of India**

Unified Prediction Model (UPM) was applied on each dataset separately from dataset 1 to dataset 17 described in Table 1.

**B    Experimental Setup and Measures**

RapidMiner Studio Educational Version 7.6.003 is used to implement UPM. The version also extends and implements algorithms designed for Weka Mining Tool.

The 10-fold cross-validation is chosen as an estimation approach to obtain a reasonable idea of classifier performance, since there's no separate test data set. This technique divide training set into 10 equal parts, 9 are applied as training set for making machine algorithm learn and 1 part is used as test set. This approach is enforced 10 times on same dataset, where every training set act as test set once.

Classification accuracy, F1 Score and Kappa are used as performance indicators. Classification accuracy is the number of correct predictions made, divided by the total number of predictions made, multiplied by 100 to turn it into a percentage. In a problem where there is a large class imbalance, a model can predict the value of the majority class for all predictions and achieve high classification accuracy. The dataset used in study suffers with this; hence accuracy measure may not be the only perfect indicator to judge the performance of model. Thus, F1 Score is also taken as another measure of performance. F1 Score is the weighted average of Precision and Recall. Therefore, this score takes both false positives and false negatives into account. Kappa is also another measure, which is used in the study. Kappa Statistics is a normalized value of agreement for chance.

**C      Application of UPM on all Datasets**

UPM was applied on each dataset separately. Model works at three levels. First level implements the concept of automated pre-processing, where raw dataset was converted to refined data. Then it was taken further at second level for classification. An ensemble method of voting with four classifiers namely Random Tree, K-Star, Simple Cart and Random Forest was used. Last 3rd Level generates rules to facilitate decision making.

## 5. Results

Unified Prediction Model was applied on 17 states datasets and results were calculated. Table 2 depicts the results obtained after applying UPM in terms of Accuracy, F1 Score and Kappa Statistics.

Figure 2, 3 and 4 represents bar chart of the results obtained. Figure 2 represents Employability Prediction Accuracy of all the 17 states, which reaches above 80% mark and average of all the states is 90.35%. Figure 3 represents Employability Prediction in terms of Precision and Recall (as represented by F1 Score) of all the 17 states, which reaches above 80% mark and average of all the states is 90.68%. Figure 4 represents Employability Prediction in terms of Kappa Score of all the 17 states, which reaches above 0.6 mark and average of all the states is 0.807. All the above results showcase that Unified Prediction Model can be applied with any dataset of Education system in India where basic parameters of aptitude, demographic, academic, cognitive and psychometric test are taken into consideration, even if data is multidimensional and unbalanced in nature.

| S.No. | State | Accuracy (%) | F1 Score | Kappa |
|-------|-------|--------------|----------|-------|
| 1 | AP | 90.6 | 90.5 | 0.812 |
| 2 | Bihar | 97.8 | 97.8 | 0.956 |
| 3 | Chattisgarh | 99 | 99.002 | 0.98 |
| 4 | Delhi | 84 | 85.38 | 0.68 |
| 5 | Gujarat | 95 | 94.88 | 0.9 |
| 6 | Haryana | 85 | 86.08 | 0.7 |
| 7 | Jharkhand | 88 | 88.54 | 0.76 |
| 8 | Karnataka | 97.8 | 97.83 | 0.956 |
| 9 | Kerala | 82.33 | 82.15 | 0.647 |
| 10 | Maharashtra | 96.78 | 96.8 | 0.936 |
| 11 | MP | 92 | 92.39 | 0.84 |
| 12 | Punjab | 82.8 | 84.13 | 0.656 |
| 13 | Rajasthan | 85 | 85.71 | 0.7 |
| 14 | Tamilnadu | 86.5 | 87.1 | 0.73 |
| 15 | UP | 83.82 | 83.6 | 0.676 |

| 16 | Uttarakhand | 89 | 89.32 | 0.78 |
| 17 | West Bengal | 98 | 98.03 | 0.962 |

**Table 2 Results of 17 States**

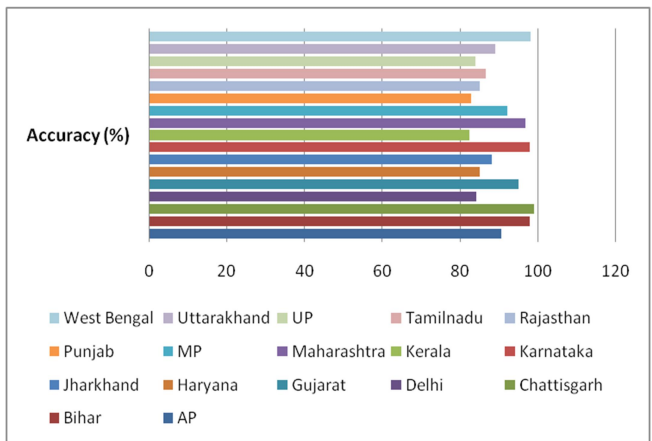

**Figure2. Employability Prediction Accuracy of 17 States**

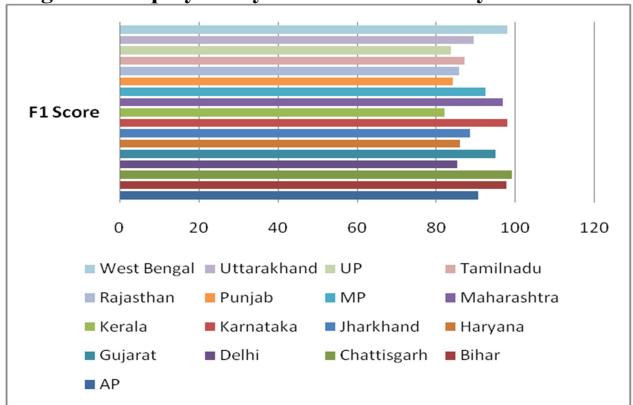

**Figure3. F1 Score of 17 States**

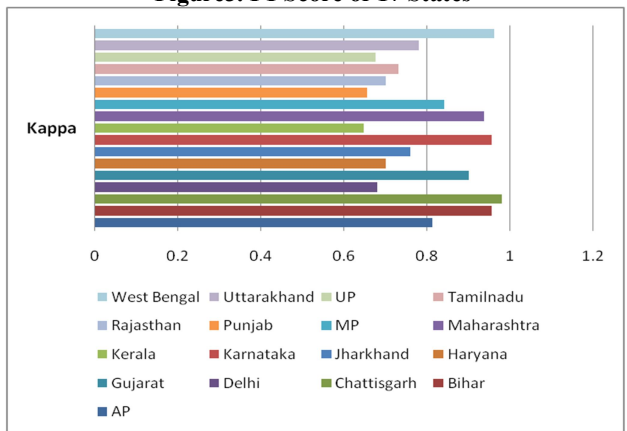

**Figure 4. Kappa Values of 17 States**

Results show case that Unified Prediction Model performed well in all the cases. Accuracy level reaches to the level of 99% in case of Chhattisgarh. Least prediction accuracy recorded in the case of Kerala with 82.33%, which is still appealing. Figure 5 showcase the Decision Tree formed in case of Chhattisgarh which depicts the highest accuracy.

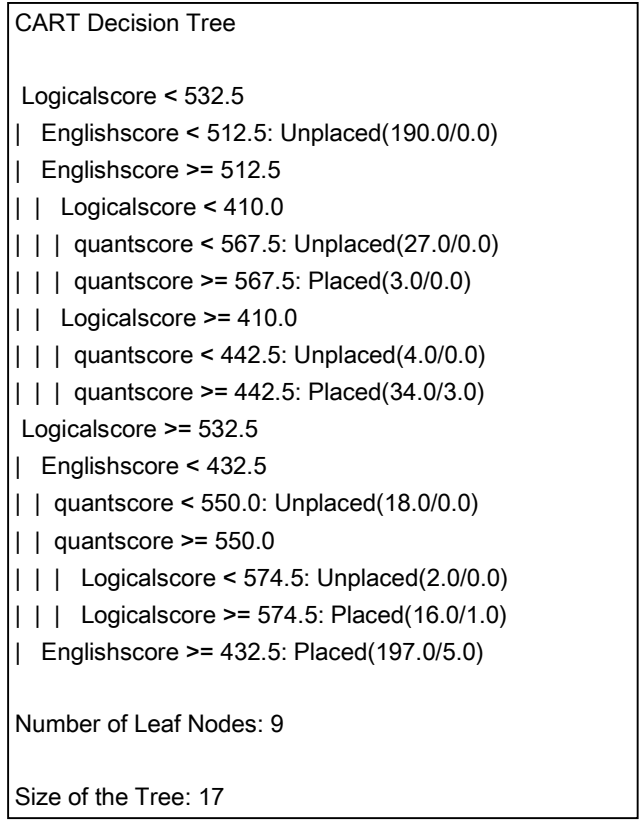

```
CART Decision Tree

 Logicalscore < 532.5
|  Englishscore < 512.5: Unplaced(190.0/0.0)
|  Englishscore >= 512.5
| |  Logicalscore < 410.0
| | |  quantscore < 567.5: Unplaced(27.0/0.0)
| | |  quantscore >= 567.5: Placed(3.0/0.0)
| |  Logicalscore >= 410.0
| | |  quantscore < 442.5: Unplaced(4.0/0.0)
| | |  quantscore >= 442.5: Placed(34.0/3.0)
 Logicalscore >= 532.5
|  Englishscore < 432.5
| |  quantscore < 550.0: Unplaced(18.0/0.0)
| |  quantscore >= 550.0
| | |  Logicalscore < 574.5: Unplaced(2.0/0.0)
| | |  Logicalscore >= 574.5: Placed(16.0/1.0)
|  Englishscore >= 432.5: Placed(197.0/5.0)

Number of Leaf Nodes: 9

Size of the Tree: 17
```

**Figure5. Cart Decision Tree for State Chhattisgarh**

It is also observed that Logical Score and Percentage, English Score and Percentage, Quant Score and Percentage plays significant role in prediction of employability in all the 17 states of India.

To further prove the homogeneity and robustness of UPM hypotheses were formed and tested with statistical parameters of t statistics, one sample test.

Hypotheses:

H0: There is not a significant difference in accuracy level of different states.

Table 3 and Table 4 represent the results of One-Sample Statistics

**Table 3: One-Sample Statistics**

|  | N | Mean | Std. Deviation | Std. Error Mean |
|---|---|---|---|---|
| Accuracy (%) | 17 | 90.24 | 6.160 | 1.494 |

**Table 4: One-Sample Test**

|  | Test Value = 90 | | | | | |
|---|---|---|---|---|---|---|
|  |  |  |  |  | 95% Confidence Interval of the Difference | |
|  | t | df | Sig. (2-tailed) | Mean Difference | Lower | Upper |
| Accuracy (%) | .157 | 16 | .877 | .235 | -2.93 | 3.40 |

P value for t statistics is 0.877, which suggests that accuracy level of all states exists at same average level which is 90% in our case. Thus model performs equally well in all the cases.

Next hypothesis is as follows
H0: There is not a significant difference in F1 Score of different states.
Table 5 and Table 6 represent the results of One-Sample Statistics

**Table 5: One-Sample Statistics**

|  | N | Mean | Std. Deviation | Std. Error Mean |
|---|---|---|---|---|
| Kappa | 17 | .804176 | .1219837 | .0295854 |

**Table 6: One-Sample Test**

|  | Test Value = .8 | | | | | |
|---|---|---|---|---|---|---|
|  |  |  |  |  | 95% Confidence Interval of the Difference | |
|  | t | df | Sig. (2-tailed) | Mean Difference | Lower | Upper |
| Kappa | .141 | 16 | .890 | .0041765 | -.058542 | .066895 |

P value for t statistics is 0.890, which suggests that accuracy level of all states exists at same average level which is 80% in our case. Thus model performs equally well in all the cases.

Next hypothesis is as follows
H0: There is not a significant difference in F1 Score Statistics of different states.
Table 7 and Table 8 represent the results of One-Sample Statistics

**Table 7: One-Sample Statistics**

|  | N | Mean | Std. Deviation | Std. Error Mean |
|---|---|---|---|---|
| F1 Score | 17 | 90.53 | 5.864 | 1.422 |

**Table 8: One-Sample Test**

|  | Test Value = 90 | | | | | |
|---|---|---|---|---|---|---|
|  |  |  |  |  | 95% Confidence Interval of the Difference | |
|  | t | df | Sig. (2-tailed) | Mean Difference | Lower | Upper |
| F1 Score | .372 | 16 | .715 | .529 | -2.49 | 3.54 |

P value for t statistics is 0.715, which suggests that accuracy level of all states exists at same average level which is 90% in our case. Thus model performs equally well in all the cases.

Therefore the entire hypotheses have been accepted and it is clear that this model can work on varied type of dataset in any university or institute, thus provides a unified generic model for predicting employability. Henceforth, UPM can be considered as robust model.

## 6. Conclusion and Future Scope

The results prove that prediction performance for students' employability can be enhanced by applying UPM. It is also proved that the model is robust in nature and can be applied on any type of unbalanced and multidimensional dataset of Education System in various contexts. Moreover, clustering applied on attributes set at pre-processing stage helps in parsimonious selection of variables and improves performance of predictive algorithms. This paper also statistically analyzes and compares the results of 17 states datasets, when applied with UPM. The results clearly depicts that Model is superior to commonly used methods of predicting students' employability in terms of universal applicability. Results prove that there is no significant difference exists in Indian Education System with respect to states as far as prediction of employability of students is concerned. Model also helps in reducing dimensionality and can further be used for domains other than education. Taking the base of UPM model, it can be tested on various types of datasets in future.

# 7. Acknowledgments

The authors gratefully acknowledge the use of services and facilities of the Faculty Research Center and Library Resources at Vivekananda Institute of Professional Studies, GGSIPU, Delhi, India.